\def\year{2021}\relax
\documentclass[letterpaper]{article} 
\usepackage{aaai21}  
\usepackage{times}  
\usepackage{helvet} 
\usepackage{courier}  
\usepackage[hyphens]{url}  
\usepackage{graphicx} 
\usepackage{csquotes}
\usepackage{amssymb}
\usepackage{rotating}
\usepackage{amsmath}
\usepackage{natbib}  
\usepackage{caption} 
\usepackage{array}
\usepackage{setspace} 
\usepackage{makecell}
\usepackage[flushleft]{threeparttable}
\usepackage{booktabs,caption}
\usepackage[toc,page]{appendix}
\usepackage{subcaption}
\usepackage{rotating}
\usepackage{multirow}
\usepackage[T1]{fontenc}

\usepackage{xcolor}

\graphicspath{ {Graphics/} }
\title{Taking sides: Public Opinion over the Israel-Palestine Conflict in 2021}

\author {
     Arsal Imtiaz,
     Danish Khan,
     Hanjia Lyu,
     Jiebo Luo \\
 }
 \affiliations {
      University of Rochester, Rochester, New York \\
     \{aimtiaz2, dkhan5, hlyu5\}@ur.rochester.edu, 
     jluo@cs.rochester.edu}

\begin{document}

\maketitle

\begin{abstract}
The Israel-Palestine Conflict, one of the most enduring conflicts in history, dates back to the start of $20^{th}$ century, with the establishment of the British Mandate in Palestine and has deeply rooted complex issues in politics, demography, religion, and other aspects, making it harder to attain resolve. To understand the conflict in 2021, we devise an observational study to aggregate stance held by English-speaking countries. We collect Twitter data using popular hashtags around and specific to the conflict portraying opinions neutral or partial to the two parties. We use different tools and methods to classify tweets into pro-Palestinian, pro-Israel, or neutral. This paper further describes the implementation of data mining methodologies to obtain insights and reason the stance held by citizens around the conflict.

\end{abstract}

\section{Introduction}

With the increasing usage of social media in the political context, Twitter maintains its stature as one of the top platforms to express political opinions~\cite{twitterintro}. It is oriented towards facilitating digital political advertising in addition to playing an active role in how candidates use them to communicate with voters~\cite{twitterintro2}. It has also allowed users to stay updated with daily news and politicians to effectively mobilize protesters~\cite{twitterintro3}.

The Israel-Palestine conflict is one of the longest-running and most controversial conflicts in the world.\footnote{\url{https://www.business-standard.com/article/international/israel-palestine-conflict-history-guide-to-world\%E2\%80\%99s-longest-running-jews-muslim-arab-war-121051300886\_1.html}} The spark between the Arabs and Jews in Palestine can be traced back to the establishment of British Mandate in 1917 during the British Occupation of Palestine~\cite{intro:1}. Since then, multiple violent confrontations took place between Jews and Arabs in Palestine, costing thousands of lives from both sides and making it an important issue for the world to address~\cite{intro:3}. 

Housing policy and design have always favored the shaping of the Jewish state and colonial domination in Israel further~\cite{intro:4}. This framework used by the Zionist Israeli regime to counter insurgent practises of Palestinians in the country sparked the conflict in 2021.\footnote{\url{https://www.vox.com/22440330/israel-palestine-gaza-airstrikes-hamas-updates-2021}} It resulted in a number of events which escalated the conflict as a Human Rights issue by the United Nations.

This study attempts to understand the response to the current Israel-Palestine conflict by English-speaking countries. The importance lies in the dissection of controversies around the conflict. The Council on Foreign Relations warned that a third intifada could break out and that renewed tensions will escalate into large-scale violence.\footnote{\url{https://www.cfr.org/global-conflict-tracker/conflict/israeli-palestinian-conflict}} Being updated about daily happenings and opinions around the conflict might untangle controversies and prevent the said intifada. The proposed methodology in the study provides an aggregated stance of English-speaking countries and gathers insights into the topics discussed by people pertaining to these countries on the grounds of their support towards Israel/Palestine.

The study makes use of modern day techniques in data mining and linguistics to study opinions based on individuality and nationality. The unresolved status of the subject instigates the need to perform a study with modern day techniques in Data Science and related fields.

\section{Related Work}

\subsection{Twitter as a Political Medium} There is an ever growing use of Twitter globally and politics plays a key role in this growth. 
With the rapid increase of the use of social media in the 2010s, analysis of data within these platforms has become one of the adapted methodologies to gather insights. Political participation in Twitter is performed both by politicians and citizens~\cite{twitterpolitics1}. The methodology mentioned in \citet{twitterpolitics2} provides insights into ``social media revolution'' and ``democratic revolution'' by studying the Maidan Revolution using Twitter data of Ukrainians. This study also aims to study public opinion in 2021 toward the Israel-Palestine conflict  to gather insights about the issue. 

\subsection{Stance Detection} 
Stance detection is the task of automatically determining from text whether the author of the text is in favor of, against, or neutral toward a proposition or target \citep{stance1}. Stance on social media data can be modeled using various online signals. These signals can be categorized into two main types, namely, (1) content signals, such as the text of the tweets,
and (2) network signals, such as the users’ connections and interactions on their social networks \citep{stance2}. In this paper, we study the former and attempt to determine stance on tweets pertaining to a bounded geographical location. In the socio-politcal domain, \citet{stance2} show that there is an orthogonal relation between sentiment and stance but studies such as \citet{stance3} show that sentiment can be used as a proxy to determine stance. In this study, we implement and compare transformer-based stance detection models and the models relying on sentiment to detect stance.

\subsection{Israel-Palestine Conflict and Social Media}
Although there is limited research on the Israel-Palestine conflict using data mining, the studies conducted on this subject have proposed efficient methods for obtaining insights. \citet{gazaconflict} collected data from social media and implemented vector autoregression to measure conflict dynamics between the Israeli Air Force (IAF) and Hamas during the Gaza conflict of 2008. \citet{related:3} also conducted a study on the Israel-Palestine conflict by using modern NLP tools, such as ArkTweetNLP and LingPipe, to analyze individual and national opinions on the Israel-Palestine conflict in the past. The study presented in this paper uses a similar methodology but focuses on the context of the Israel-Palestine conflict in 2021 and considers a larger dataset. We also make use of visualization tools and methods to obtain inferences regarding support.

\section{Problem Statement}

The objective of this paper is to study stance in English-speaking countries around the Israel-Palestine conflict in 2021. It further attempts to reason support towards Israel/Palestine in English-speaking countries with the help of techniques and methods in Natural Language Processing.

\section{Methodology}
\begin{figure}[htp]
    \centering
    \includegraphics[width=\linewidth]{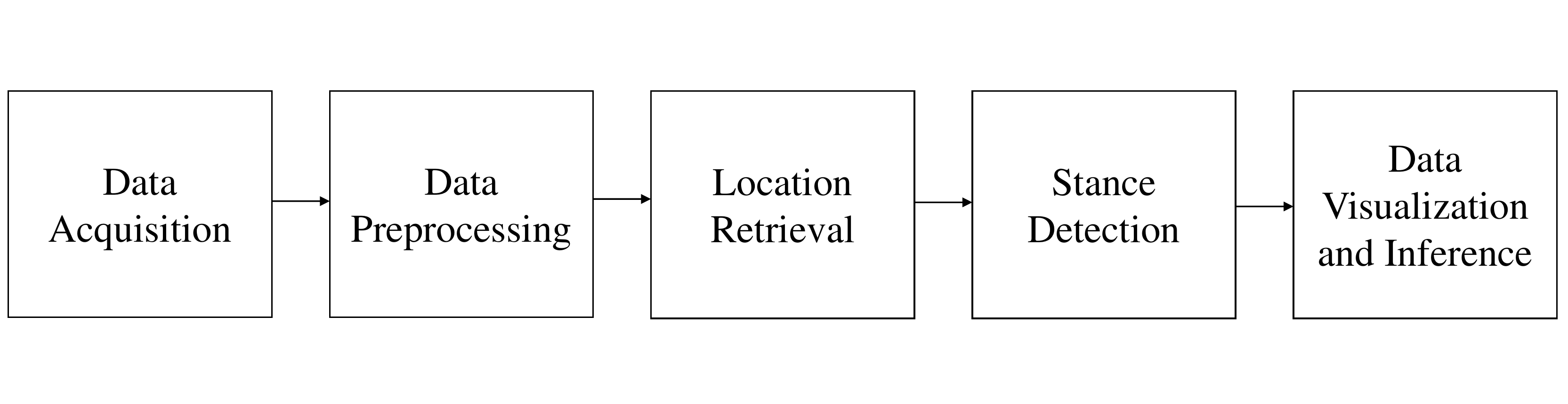}
    \caption{Pipeline of the analysis.}
    \label{fig:block diagram.png}
\end{figure}

This section of the paper discusses the pipeline for the study. As shown in Figure~\ref{fig:block diagram.png}, the data acquired is passed through multiple stages presented in the pipeline to obtain semantic understanding and insights around the conflict. The stages of the pipeline are explained in detail in the subsections below.

\subsection{Data Acquisition}
Social media platforms such as Facebook, Twitter, Reddit, \textit{etc}., can be used to convey different forms of messages, and hence, their role in propagating political information is not surprising~\cite{dataacquisition1}. The findings of \citet{dataacquisition2} suggested that ``Twitter users gain in political knowledge by the virtue of their use of that medium'', which supports the argument regarding the political nature of Twitter. 

For this study, we have considered Twitter data because of its speed, immediacy, ease of use via hashtags, and its global reach~\cite{dataacquisition3}. Twitter API\footnote{\url{https://developer.twitter.com/en/docs/twitter-api}} is used to scrape english tweets around the timeline in the format shown in Table~\ref{Table1}. The API, under academic access allows us to scrape tweets at a faster rate with a higher number of API calls, and provides more information about each tweet and its associated user.

\begin{table}[h]
\begin{center}
\small
\begin{tabular}{|l|l|}
\hline
\textbf{Field} & \textbf{Description}\\
\hline
text  & Tweet text content \\ 
language & Tweet language \\  
tweet\_id & Unique tweet ID \\
source & Device used for posting the tweet\\
conversation\_id & Unique ID for every conversation\\
retweet\_count & Number of retweets\\
reply\_count & Number of replies to the tweet\\
like\_count & Number of likes of the tweet\\
quote\_count & Number of times the tweet has been quoted\\
author\_id & Unique user ID\\
created\_at & Datetime of the posted tweet\\
\hline
\end{tabular}
\caption{Summary of the tweet fields.}
\label{Table1}
\end{center}
\end{table}

\begin{figure*}[t]
    \centering
    \includegraphics[width=\linewidth]{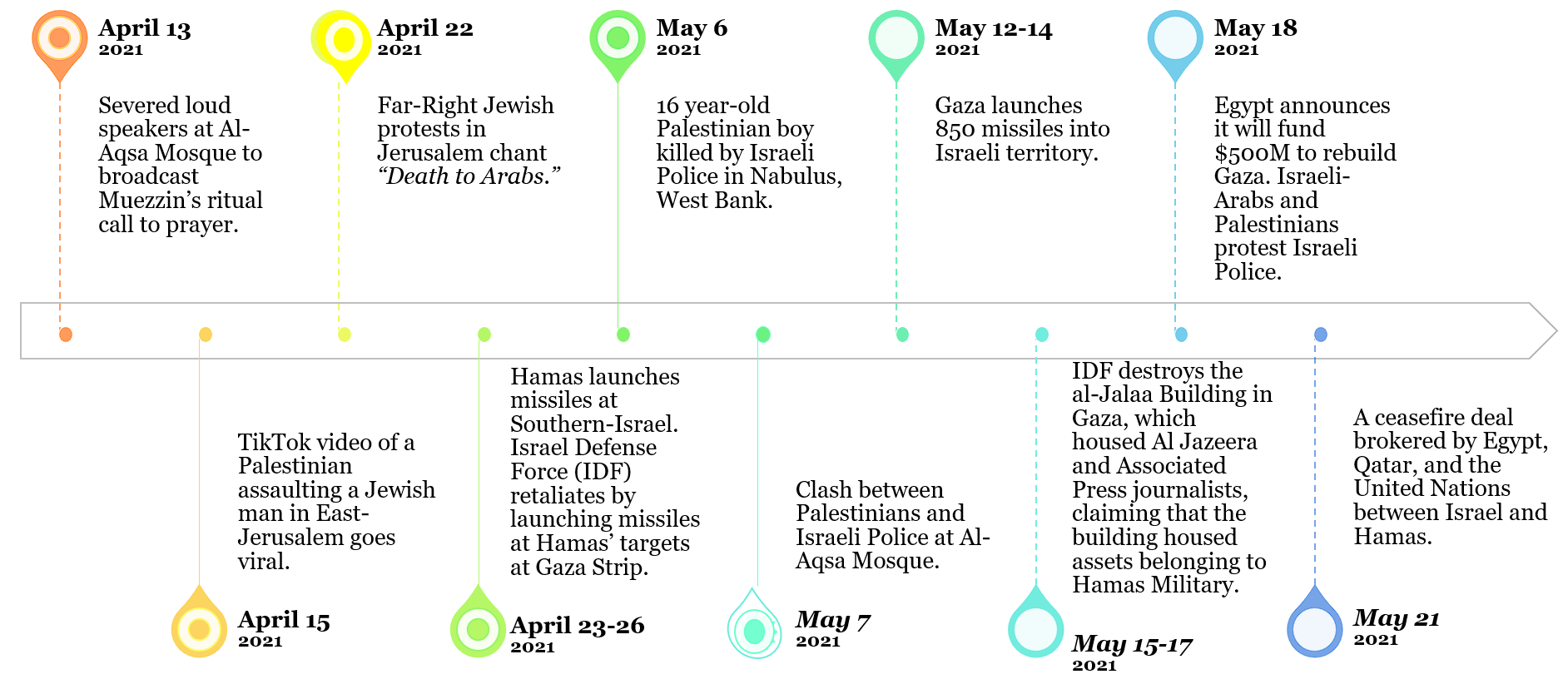}
    \caption{The timeline of the Israel-Palestine Conflict in 2021.}
    \label{fig:timeline}
\end{figure*}

Data acquisition was carried out in primarily two phases with respect to the timeline shown in Figure~\ref{fig:timeline}. First, a list of popular hashtags regarding the conflict is curated. A small sample of tweets of roughly 1,500-2,000 tweets per day are extracted over the entire timeline of the conflict. The hashtags specific to the study such as \textit{\#freepalestine, \#antisemitism, \#istandwithpalestine, \#istandwithisrael, \#jerusalemfightsback, etc.} are used. The hashtags acquired in the first phase are then used as parameters of acquiring the larger dataset. 

The API configuration is set to scrape 200 tweets every ten minutes of the entire timeline. A total of 493,877 tweets ranging from April $13^{th}$, 2021 to May $23^{rd}$, 2021 are collected.

\subsection{Data Preprocessing}
 
Data preprocessing is an essential step to formulate the data in a form suitable for stance detection. In this stage, we will discuss data cleaning, demojizing emoticons, part-of-speech subject estimation, \textit{etc}.


Data cleaning for our data includes removal of stopwords, hyperlinks, HTML escape sequences, white-spaces, \textit{etc}. with regular expressions and NLTK libraries in Python. Hashtags on the other hand are retained to preserve meaning in the tweets~\cite{preprocess2}. For instance, ``Save children in \#Gaza'' would have lost its context if "\#Gaza" is removed entirely from the tweet. 

\citet{preprocess3} stressed on the possibility that users may emphasize emotions by the use of emoticons along with co-occurring emotional words or complement the emotional words presented in the tweet. Hence, we demojize emoticons in the tweet to retain meaning. Emoji,\footnote{\url{https://github.com/carpedm20/emoji/}} a python library is used to convert emoticons to text such as happy, sad, angry, excited, \textit{etc}.

\citet{preprocess4} recognised part-of-speech tagging as one of the most fundamental parts in a linguistic pipeline that often leads to a richer linguistic analysis. Therefore, we use SpaCy part-of-speech tagging across our data to estimate the subject of the tweet.

\subsection{Location Retrieval}
Only an estimated 2\% of the tweets are geo-tagged with latitude and longitude coordinates, and there is a need for methods to assign geo-locations to each tweet and its associated user~\cite{location1}. Primarily we intend to extract user location by referring to the ``geo'' field of each tweet. If the tweet is geo-tagged, it is stored as a location in the location field. For the situations where the tweet is not geo-tagged, the user's location, as mentioned in their bio is fetched using user fields from the API's .json file. If the user has not mentioned their location in their bio, their location is stored as {\tt NULL}. Python's GeoPy library\footnote{\url{https://geopy.readthedocs.io/en/stable/}} is used to process the locations to a universal country format for the entirety of data.



\subsection{Stance Detection}

Twitter provides a great corpus of text for stance detection and opinion mining based on its sheer usage~\cite{SA1}. In this study, we have implemented stance detection to classify tweets into three classes - pro-Palestine, pro-Israel, and neutral via two different approaches. The first approach uses pre-trained sentiment analysis tools such as VADER~\citep{vader}, LIWC,\footnote{https://www.liwc.app/} PySentimiento~\citep{pysentimiento} as a part of the model. These tools provide an array of features on the data, and are clubbed with logistic regression/multi-layer feedforward neural network for text classification.

The structure of the multi-layer feedforward neural network is described in Table~\ref{NeuralNet}. The input of the network is configured to the outputs of VADER, LIWC and PySentimiento. There exists three hidden layers with 512, 256 and 64 neurons respectively. The output layer of the model has 3 neurons computing the probability for each class using a softmax function. The model is configured to have a learning rate of 0.001, batch size of 20, categorical cross entropy as the loss function, Adam as the optimizer and is trained for 500 epochs.

The second approach attempts to solve the problem of stance detection using pre-trained transformer-based models. We have fine-tuned an XLNet~\citep{xlnet} model on our corpus of tweets to determine stance. The pre-trained model used for this study is \textit{xlnet-base-cased}. The model is configured to have three outputs for stance detection and is trained for 10 epochs with a batch size of 32.

For training and testing models, we have manually annotated 979 tweets into three classes - pro-Palestine (n=510), pro-Israel (n=290) and neutral (n=179). The labeled data is then split to training and testing sets in the ratio of 3:1.


\begin{table*}[t]
\centering
\tiny

\begin{tabular}{|c|c|c|c|c|c|c|c|c|c|c|}

\hline
\multirow{2}{*}{\bfseries Model} & \multirow{2}{*}{\bfseries Accuracy} &\multicolumn{3}{|c}{\bfseries Precision}  & \multicolumn{3}{|c}{\bfseries Recall} & \multicolumn{3}{|c}{\bfseries F1-Score } \vline\\ \cline{3-11} 
\multicolumn{1}{|c}{} & \multicolumn{1}{|c}{} &\multicolumn{1}{|c}{\bfseries Pro-Palestine} & \multicolumn{1}{|c}{\bfseries Neutral} & \multicolumn{1}{|c}{\bfseries Pro-Israel}&\multicolumn{1}{|c}{\bfseries Pro-Palestine} & \multicolumn{1}{|c}{\bfseries Neutral} & \multicolumn{1}{|c}{\bfseries Pro-Israel}&\multicolumn{1}{|c}{\bfseries Pro-Palestine} & \multicolumn{1}{|c}{\bfseries Neutral} & \multicolumn{1}{|c}{\bfseries Pro-Israel} \vline \\ \hline 
     Vader with Logistic Regression & 0.63 & 0.70 & 0.44 & 0.58 & 0.69 & 0.19 & 0.81 &0.70 & 0.27 & 0.68   \\ \hline
     Vader with Neural Network & 0.63 & 0.71 & 0.32 & 0.57 & 0.77 & 0.17 & 0.64 &0.74 & 0.22 & 0.60   \\ \hline
     LIWC with Logistic Regression & 0.66 & 0.76 & 0.28 & 0.72 & 0.76 & 0.28 & 0.73 & 0.76 & 0.28 & 0.72   \\ \hline
     LIWC with Neural Network & 0.60 & 0.61 & 0.39 & 0.63 & 0.78 & 0.19 & 0.54 & 0.68 & 0.25 & 0.58   \\ \hline
     PySentimiento with Logistic Regression & 0.65 & 0.74 & 0.28 & 0.69 & 0.74 & 0.26 & 0.71 &0.74 & 0.27 & 0.70   \\ \hline
     PySentimiento with Neural Network & 0.67 & 0.65 & 0.62 & 0.71 & 0.89 & 0.19 & 0.60 &0.76 & 0.20 & 0.65   \\ \hline
     \textbf{XLNet} & \textbf{0.82} & \textbf{0.84} & \textbf{0.59} & \textbf{0.91} & \textbf{0.82} & \textbf{0.56} & \textbf{0.97} &\textbf{0.83} & \textbf{0.57} & \textbf{0.94}  \\
     \hline
\end{tabular}
\caption{Model comparison: accuracy, precision, recall and F1-score.}
\label{ModelComp}
\end{table*}



\begin{table}[t]
\begin{center}
\small
\begin{tabular}{|l|l|l|}
\hline 
\textbf{Layer} & \textbf{Activation Function} & \textbf{Output Shape}\\
\hline
Dense & relu & (None, 512)  \\ \hline
Dense & relu & (None, 256)  \\ \hline
Dense & relu & (None, 64)  \\ \hline
Dense & softmax & (None, 3)  \\ \hline

\hline
\end{tabular}
\caption{Multi-Layer Feedforward Neural Network Structure.}
\label{NeuralNet}
\end{center}
\end{table}


The models are passed through the testing set and their accuracy, precision, recall,  F1-score  are reported in Table~\ref{ModelComp}. Our results confirm the findings of \citet{stance2} that stated the relation between sentiment and stance as orthogonal.

\subsection{Data Visualization and Inferences}

People are generally more engaged when information is represented in a space and grid form in comparison to representation in the form of lists and statistics. Visualizing opinions in a political context may enhance intentional and deliberate cognition, and interpretations, whilst mitigating framing effects and subconscious processes~\cite{visual1}. In this study, we have made use of \textit{Tableau} and \textit{Wordart}\footnote{\url{https://wordart.com/}} to generate bar plots and wordclouds.




\section{Results}

This section provides statistical information about the data in Table~\ref{Table4} and attempts to reason stances held by English-speaking countries.  Given the performances reported in Table~\ref{ModelComp}, all the stance-related statistics in Table~\ref{Table3} are based on the stance determined by the XLNet model.
\begin{table}[t]
\begin{center}
\small
\begin{tabular}{|l|l|}
\hline
\textbf{Tweet Metric} & \textbf{Value}\\
\hline
Average number of words per tweet & 21.63 \\ 
Median number of words per tweet& 21.63 \\  
Minimum number of words & 1 \\
Maximum number of words& 133 \\
Standard deviation of the number of words & 12.702 \\
Total number of tweets & 493,877\\
Number of pro-Israel tweets & 14,982\\
Number of neutral tweets & 250,109\\
Number of pro-Palestine tweets & 228,813\\
\hline
\end{tabular}
\caption{Descriptive statistics of the tweets.}
\label{Table3}
\end{center}
\end{table}





\begin{figure}[htp]
    \centering
    \includegraphics[width=\linewidth]{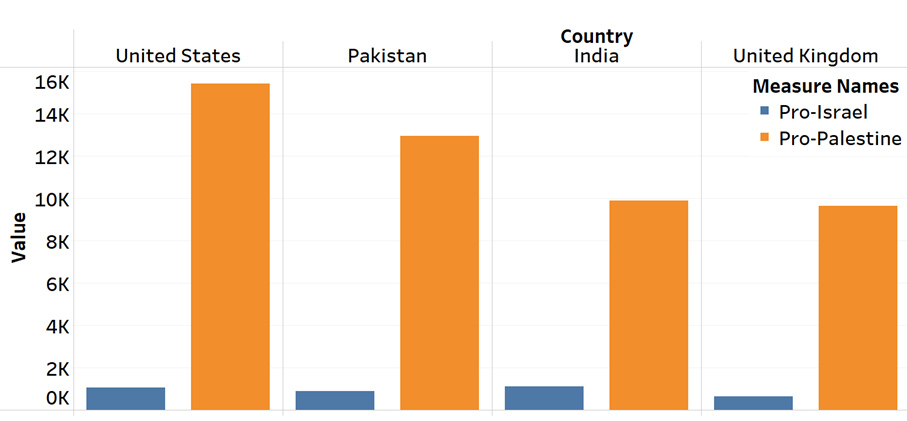}
    \caption{Number of pro-Israel and pro-Palestine tweets of the top four countries.}
    \label{fig:barplot}
\end{figure}

We have set a threshold of 10,000 tweets per country to consider their stance in our study. Table~\ref{Table4} provides the total number of tweets by the four most actively opinionated English-speaking countries in this conflict. Figure~\ref{fig:barplot} shows a clear support for Palestine in the discussed conflict for these countries.
\begin{table}[h!]
\small
\begin{center}
\begin{tabular}{|l|l|}
\hline
\textbf{Country} & \textbf{No. of Tweets}\\
\hline
United States     &        38,418 \\
Pakistan          &        23,346 \\
United Kingdom    &        22,611 \\
India             &        19,543 \\
\hline
\end{tabular}
\caption{Ranking of countries in terms of expressing opinions around the recent conflict.}
\label{Table4}
\end{center}
\end{table}

Wordclouds are generated to visualize topics that Palestine/Israel supporters are talking about. We attempt to reason their opinions and juxtapose the events that took place. The highly frequent words attained from the wordclouds can be queried in the data to understand and reason support among pro-Israeli/pro-Palestinian supporters. The inferences attained from pro-Palestinian tweets can be assumed to sound the opinion of the masses but the inference attainable from pro-Israeli tweets poses a higher possibility of opinions to be specific and deep rooted in culture. We consider the case of United States as an example in this study. The same methodology can be applied to other countries to attain inferences as well.

\begin{figure}[t]
    \centering
    \includegraphics[width=\columnwidth]{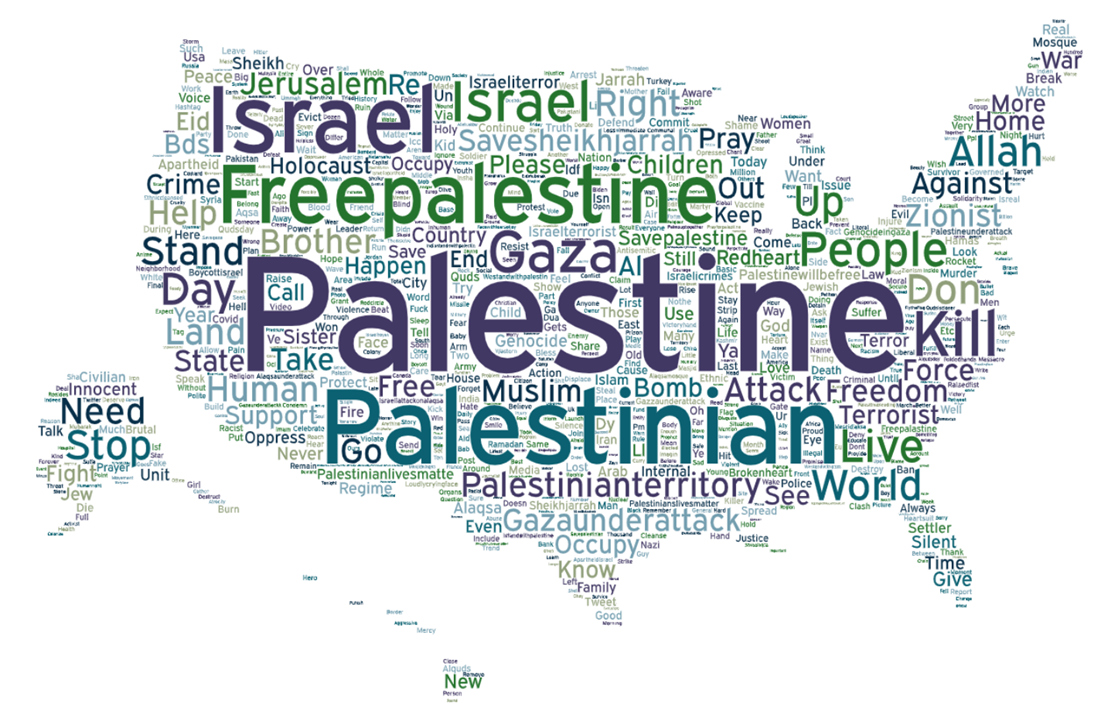}
    \caption{Pro-Palestinian tweets in the United States.}
    \label{fig:wordcloud2}
\end{figure}

With the United States being the major contributor of the data, the finding presents some interesting insights. According to Figure~\ref{fig:wordcloud2},  \textit{freepalestine, Gaza, Savesheikhjarrah}, \textit{etc}. are some of the frequent words that have occurred amongst Palestinian supporters in the United States. On querying tweets with the words obtained above, there is a clear expression of contempt toward the events within the timeline. For example, there have been tweets that depict the condemnation of children's death in Gaza.

Upon further querying and studying tweets, a number of Palestinian supporters expressed discontentment toward the President and the Vice President of the United States. The supporters were not able to accept with the fact that US President approved sale of weapons worth \$735M to Israel.\footnote{\url{https://www.washingtonpost.com/politics/2021/05/17/power-up-biden-administration-approves-735-million-weapons-sale-israel-raising-red-flags-some-house-democrats/}} The sale of weapons to Israel was a counterproductive move in the efforts of waning and stopping the conflict.

\begin{figure}[htp]
    \centering
    \includegraphics[width=\linewidth, trim = {0.15cm 0 0 0},clip]{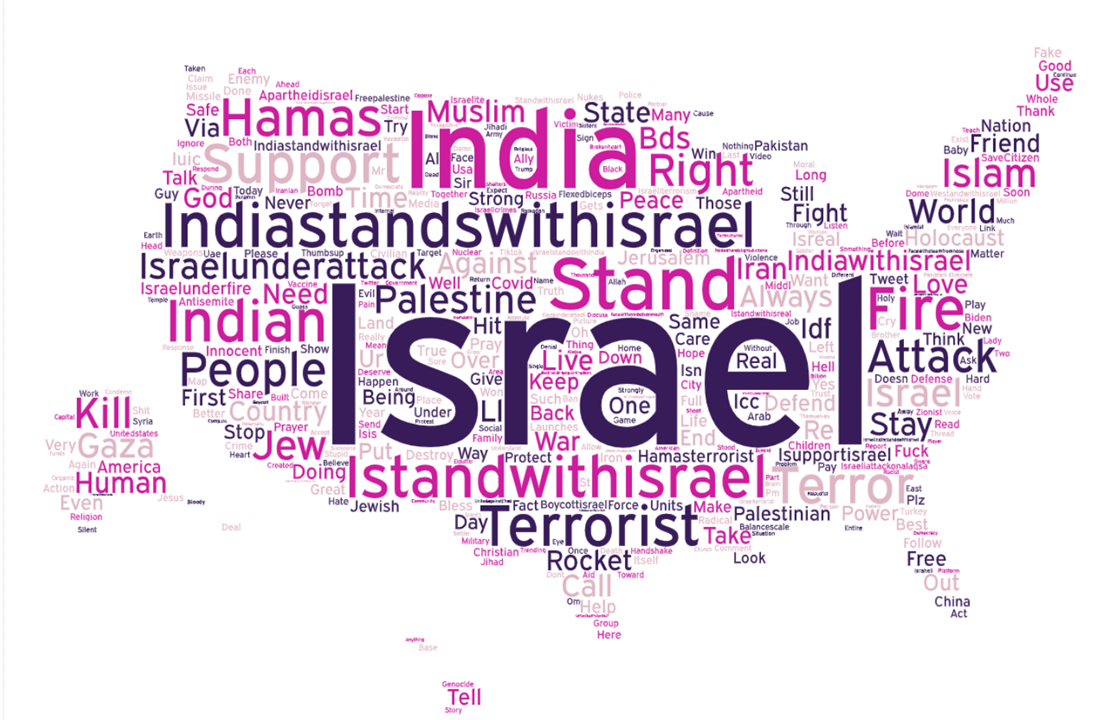}
    \caption{Pro-Israeli tweets in the United States.}
    \label{fig:wordcloud1}
\end{figure}

Similarly, opinions around Israeli supporters can also be drawn with this methodology. Israeli supporters expressed opinions about topics in the timeline that are generally associated with negative connotations. Figure~\ref{fig:wordcloud1} shows a few topics - \textit{Hamas, terrorist, rocket, attack}, \textit{etc.} that are frequent among pro-Israeli tweets. On querying, there is a strong possibility that the condemnation rises from the hate of terrorism - \textit{Hamas} in this case. 

\section{Conclusion}

In this study, we propose a methodology to present insights into the public opinion toward the conflict at hand. It possesses the ability to obtain inferences about stance toward Israel and Palestine in accordance to real time events and evidence with respect to individual nations. With the acquisition of data in such a distinct domain, we present data in visuals encompassing the trends of support in the conflict with events and occurrences around the conflict in the United States. Our study supports the claim by \citet{results1} of using Twitter more as an outlet for expressing discontent than as a medium for negative campaigning. However, the basis of our study on English-speaking countries excludes the valued opinions of non-English speaking countries, especially near Israel-Palestine which would have an impact on the study. Inclusion of the said tweets would have provided greater clarity on the situation based on multi-lingual stance detection.

From a technical standpoint, we find that the transformer-based model - XLNet outperforms off-the-shelf hybrid implementations of sentiment analysis tools in political domains specific to conflicts between parties, which is in line with the findings of \citet{bestvater} that political opinions are highly complex and sentiment analysis methods may not capture stance. There exists a class imbalance issue in the data, with pro-Israel tweets being very few in comparison to pro-Palestine tweets, which can be addressed in future work. Multi-lingual models can be considered in the future to take other popular languages such as Arabic, Hebrew, \textit{etc}. into account. This would help us process and understand data around the conflict from a global perspective.

\section{Broader Impact and Ethics Statement}

The study possesses the potential of gathering numerous insights on a broader scale. Considering specific time frames around heated events provides detailed voicing of people irrespective of their involvement in the conflict. It provides a mechanism into understanding why people support Israel/Palestine in the conflict. However, patterns and insights attained from the conflict can also be used in a detrimental way to filter out tweets by organizations or governments. 

According to the Twitter Developer Agreement and Policy, the organization prohibits the use of data in any way that would breach user privacy. For non-commercial purposes such as academic research, Twitter provides some liberty in distribution of content but limits the scope to distributing User IDs, Tweet IDs and Direct-Message IDs only. Making the data with a broader array of information such as tweets, geo-data and other personal information puts the licensee and the organization in jeopardy for potential breach of privacy. Sensitive Twitter data is collected in the study for gathering insights at an aggregate level without disclosing any personal or sensitive information about users.

\bibliography{name}
\bibstyle{aaai}

\end{document}